\documentstyle[amssymb,12pt]{article}
\makeatother
\begin{document}

\begin{center}
{\Large {\bf The stopping of \ swift protons in matter and its implication
for astrophysical fusion reactions}}

{\Large {\bf \ }}

{\Large \ C.A. Bertulani}$^{\left( a,b\right) (*)}${\Large \ and D.T. de Paula}$%
^{(a)}$\footnote{%
E-mails: bertu@if.ufrj.br, dani@if.ufrj.br}

$^{(a)}$Instituto de F\'{i}sica, Universidade Federal do Rio de Janeiro,
21945-970 Rio de Janeiro, RJ, Brazil\\[0pt]
$^{(b)}$Brookhaven National Laboratory, Physics Department, Upton, NY
11973-5000, USA
\end{center}

\begin{quotation}
The velocity dependence of the stopping power of swift protons and deuterons
in low energy collisions is investigated. At low projectile energies the
stopping is mainly due to nuclear stopping and charge exchange of the
electron. The second mechanism dominates after E$_{p}\geq 200$ eV. A
dynamical treatment of the charge exchange mechanism based on two-center
electronic wavefunctions yields very transparent results for the exchange
probability. \ We predict that the stopping cross sections vary
approximately \ as v$_{p}^{1.35}$ for projectile protons on hydrogen targets
in the 1 keV energy region.
\end{quotation}

\bigskip

Nuclear fusion reactions proceed in stars at extremely low energies, e.g.,
of the order of $10$ keV in our sun \cite{Cla68,RR88}. At such low energies
it is extremely difficult to measure the cross sections for charged
particles at laboratory conditions due to the large Coulomb barrier. One
often uses a  theoretical model to extrapolate the experimental data to the
low-energy region. Such extrapolations are sometimes far from reliable, due
to unknown features of the low-energy region. E.g., there might exist
unknown resonances along the extrapolation, or even some simple effect which
one was not aware of \ before. One of these effects is the laboratory atomic
screening of fusion reactions \cite{ALR87,SR95}. It is well known that the
laboratory measurements of low energy fusion reactions are strongly
influenced by the presence of the atomic electrons. This effect has to be
corrected for in order to relate the fusion cross sections measured in the
laboratory with those at the stellar environment. Another screening effect,
\ arising from free electrons in the stellar plasma, will not be treated
here. For about one decade, until 1996, one observed a large discrepancy
between the experimental data and the best models available to treat the
screening effect. The simplest (and perhaps the best of these models), the
so-called adiabatic model, predicts that as the projectile nucleus
penetrates the electronic cloud of the target the electrons become more
bound and the projectile energy increases by energy conservation. Since the
fusion cross sections increase strongly with the projectile's energy, this
tiny amount of energy gain (of order of 10-100 eV) leads to a large effect
on the measured cross sections. However, in order to explain the
experimental data, it is necessary an extra-amount of energy - about twice
the value obtained by the adiabatic model. This is puzzling, since more
refined dynamical models, e.g., time-dependent Hartree-Fock \cite{SKLS93}, \
include electronic excitation and thus yield a screening energy which is
smaller than that obtained with the adiabatic model.

This problem was apparently solved in 1996 by Langanke and collaborators 
\cite{LSBR96} and by Bang and collaborators \cite{BFMH96}, who observed that
the experimental data for $^{3}He(d,$ p$)^{4}He$ - the reaction for which
the screening effect was best studied - was probably obtained with a
erroneous extrapolation of the stopping power for deuterons in helium
targets to the low energy regime. The fusion reaction occurs at a point
inside the target after the projectile has slowed down by interactions with
the atomic targets. In the experimental analysis one needs to correct for
this energy loss in order to assign the right projectile energy value for
that reaction. These corrections were usually based on the Andersen-Ziegler
table of stopping power of low energy particles \cite{AZ77}. Due to the lack
of experimental information on the stopping power at the extreme low
projectile energies needed for astrophysical purposes, the Anderson-Ziegler
tabulation was extrapolated to the required energy; another example of a
dangerous extrapolation procedure. In fact, Golser and Semrad \cite{GS91}
observed a strong departure of their experimental data from the
extrapolations based on the Andersen-Ziegler tables for the stopping of low
energy protons on helium targets. Grande and Schwietz \cite{GS93} performed
a dynamical calculation of the energy dependence of the stopping power for
this system and confirmed that the extrapolation procedure cannot be
extended to the very low energies. Whereas at higher energies the stopping
is mainly due to the ionization of the target electrons, at the
astrophysical energies it is mainly due to charge-exchange between the
target and the projectile. Refs. \cite{LSBR96} and \cite{BFMH96} use these
arguments to explain the long standing discrepancy between theory and
experiment for the low energy dependence of the reaction $^{3}He(d,$ p$%
)^{4}He$. Other reactions of astrophysical interest (e.g., those listed in
by Rolfs and collaborators \cite{ALR87,SR95}) should also be corrected for
this effect.

In this work we address the problem of the stopping of very low energy ions
in matter. To simplify matters, we study the system $p+H$, which is the
simplest one can think of. It displays important features of the stopping
power and has the advantage of allowing a very simple solution.

Our approach is based on the solution of the time-dependent Schr\"{o}dinger
equation for the electron in a dynamical two-center field. The static
two-center $p+H$ system has been solved by Edward Teller in 1930 \cite{Te30}%
. He showed that as the distance between the protons decreases the hydrogen
orbitals split into two or more orbitals, depending on its degeneracy in the
two-center system. Analogous problems are well known in quantum systems \cite
{Me97}. For example, take two identical potential wells at a certain
distance. For large distances the states in one well are degenerated with
the states in the other potential well. As they approach this degeneracy is
lifted due to the influence of barrier tunneling. Thus, the lowest energy
state of hydrogen, $1$s, splits into the 1s$\sigma $ and the 2p$\sigma $
states as the protons approach each other. The 1s$\sigma $ state is space
symmetrical, while the 2p$\sigma $ state is antisymmetric. As the proton
separation distance decreases their respective energies decrease. At $%
R\simeq 1$ \AA\ the energy of the 2p$\sigma $ state starts to increase
again, while the energy of the 1s$\sigma $ state continues to decrease. For
proton distances much smaller than $1$ \AA\ the 1s$\sigma $ and the 2p$%
\sigma $ energies correspond to those of the first and second states of the
He atom, respectively \cite{Te30}.

Let us now consider the dynamical case. The full time-dependent wavefunction
for the system can be expanded in terms of two-center states, $\phi _{n}(t)$%
, governed by the Schr\"{o}dinger's equation

\begin{equation}
\left[ H_{0}+V_{p}\left( t\right) \right] \phi _{n}(t)=E_{n}\left( t\right)
\phi _{n}(t)\;,{\rm  \ \ \ \ \ with \ \ \ }H_{0}=\widehat{{\bf p}}%
_{e}^{2}/2m_{e}+V_{T},  \label{schro}
\end{equation}
where $V_{p}\left( t\right) =-e^{2}/\left| {\bf r}+{\bf R}/2\right| $ is the
electron-projectile proton interaction potential and $V_{T}=-e^{2}/\left| 
{\bf r}-{\bf R}/2\right| $  is the electron-target proton interaction for a
proton-proton separation distance $R(t)$. Note that in our formulation the
two-center wave functions also depend on time, as well as their energies $%
E_{n}\left( t\right) $. The full electronic wavefunction is obtained by a
sum over all orthonormal two-center states

\begin{equation}
\left| \Psi \left( t\right) \right\rangle =\sum_{n}a_{n}\left( t\right)
\left| \phi _{n}\left( t\right) \right\rangle ,\;\;\;\;{\rm with\ \ }\int
d^{3}r\phi _{n}\left( t\right) \phi _{m}\left( t\right) =\delta _{nm}\;.
\label{expand}
\end{equation}

Inserting this expansion in eq. (\ref{schro}) we obtain

\begin{equation}
i\hbar \frac{d}{dt}a_{m}\left( t\right) =E_{m}\left( t\right) a_{m}\left(
t\right) -i\hbar \sum_{n}a_{n}\left( t\right) \left\langle m\right| \frac{d}{%
dt}\left| n\right\rangle \;.
\end{equation}

Using (\ref{schro}) one can easily show that, for $m\neq n$,

\begin{equation}
\left\langle m\right| \frac{d}{dt}\left| n\right\rangle =\frac{\left\langle
m\right| dV_{p}/dt\left| n\right\rangle }{E_{n}\left( t\right) -E_{m}\left(
t\right) },\;\ \ \ \ \ \ (m\neq n).  \label{2ndpot}
\end{equation}

Moreover, using the second relation of eq. (\ref{expand}), one can show that 
$\left\langle m\right| \frac{d}{dt}\left| m\right\rangle =0$, if $\left|
m\right\rangle $ is real. This indeed will be our case. Our basis, $\left|
n\left( t\right) \right\rangle $, is formed by two-center states at a given
time $t$, i.e., a given proton separation distance, $R$. These wavefunctions
are real. Thus, the final coupled-channels equation for the two-center
problem is given by

\begin{equation}
i\hbar \frac{d}{dt}a_{m}\left( t\right) =E_{m}\left( t\right) a_{m}\left(
t\right) -i\hbar \sum_{m\neq n}a_{n}\left( t\right) \frac{\left\langle
m\right| dV_{p}/dt\left| n\right\rangle }{E_{n}\left( t\right) -E_{m}\left(
t\right) }.  \label{tdschro2}
\end{equation}

At very low proton energies $\left( E_{p}\lesssim 1{\rm  keV}\right) $ it
is fair to assume that only the low-lying states are involved in the
electronic dynamics. Only at proton energies of order of 25 keV the proton
velocity is comparable to the electron velocity, v$_{e}\simeq \alpha c$.
Thus, the evolution of the system is almost adiabatic at $E_{p}\lesssim 10$
keV. The higher states require too much excitation energy and belong to
different degeneracy multiplets. The initial electronic wavefunction is a
clear superposition of 1s$\sigma $ and 2p$\sigma $ two-center states. One
thus expects that only these states are relevant for the calculation. In
fact, at these energies the population of the 2p atomic state in charge
exchange is much less than the population of the 1s atomic state. These
assumptions are well supported by the calculations of Grande and Schwietz 
\cite{GS93}, who have used a dynamical approach based on target-centered
wavefunctions. In their approach one has to include a great amount of
target-centered states in order to represent well the strong distortion of
the wavefunction as the projectile closes in the target. We also have
assumed that the proton follows a classical trajectory determined by an
impact parameter $b$.

Eq. (\ref{tdschro2}) does not look like the usual form of coupled-channels
equations in the theory of the time-dependent Schr\"{o}dinger equation. But
we can put it in such form by rewriting the equation as

\begin{equation}
i\hbar \frac{d}{dt}\left( 
\begin{array}{c}
a_{+} \\ 
a_{-}
\end{array}
\right) =\left( 
\begin{array}{cc}
V_{+}+E_{0} & iW \\ 
iW & V_{-}+E_{0}
\end{array}
\right) \left( 
\begin{array}{c}
a_{+} \\ 
a_{-}
\end{array}
\right) \;,  \label{coupled}
\end{equation}
where the indices $+$ and $-$ refer to the $1$s$\sigma $ and $2$p$\sigma $
states, respectively, $E_{0}=-13.6$ eV, $V_{\pm }\left( t\right) =E_{\pm
}(t)-E_{0}$, and 
\begin{equation}
W\left( t\right) =\hbar \frac{\left\langle \Psi _{+}\right| dV_{p}/dt\left|
\Psi _{-}\right\rangle }{E_{+}\left( t\right) -E_{-}\left( t\right) }\equiv
\hbar \frac{\left\langle \Psi _{1{\rm s}\sigma }\left( t\right) \right|
dV_{p}/dt\left| \Psi _{2{\rm p}\sigma }\left( t\right) \right\rangle }{E_{1%
{\rm s}\sigma }\left( t\right) -E_{2{\rm p}\sigma }\left( t\right) }\;.
\label{potent}
\end{equation}
In this form, the potentials $V_{\pm }\left( t\right) $ and $W\left(
t\right) $ act like potentials in  usual coupled-channels equations. We use
the formalism of Teller \cite{Te30} to calculate the wavefunctions $\Psi
_{\pm }\left( R\right) $ at different inter-proton distances, $R(t)$,
corresponding to a particular time $t$. The static Schr\"{o}dinger equation
is solved in elliptical coordinates. This yields two coupled differential
equations which can be solved by expanding the solutions in Taylor series. A
set of recurrence relations is obtained for the expansion coefficients when
the boundary conditions are used. The energies $E_{1{\rm s}\sigma }\left(
R\right) $ and $E_{2{\rm p}\sigma }\left( R\right) $ are obtained by
adjusting the constant which separates the two coupled equations \cite{Te30}
to its correct matching value.

When $t\longrightarrow \pm \infty $, $V_{\pm }\longrightarrow 0$ and $\
W\longrightarrow 0$. The initial state, an electron localized in the target
can be written in terms of the degenerate symmetric, $\Psi _{+}=$ $\Psi _{1%
{\rm s}\sigma }$, and anti-symmetric, $\Psi _{-}=$ $\Psi _{2{\rm p}\sigma }
$, states:

\begin{equation}
\Phi _{T}=\frac{1}{\sqrt{2}}\left( \Psi _{+}+\Psi _{-}\right) ,\;\;\;\;\;\;%
{\rm at}\;\;\;t\longrightarrow -\infty \;,  \label{istate}
\end{equation}
where both $\Phi _{T}$, and $\Psi _{\pm }$ are normalized wavefunctions. If
the electron is localized in the projectile, the wavefunction $\Phi
_{p}=\left( \Psi _{+}-\Psi _{-}\right) /\sqrt{2},\;$when $t\longrightarrow
-\infty $ is used. We will consider only the condition of eq.$\left( {\rm %
\ref{istate}}\right) ,$ namely, an electron localized at the target at $%
t\longrightarrow -\infty $. \ These relations are well known quantum
mechanical results; the asymptotic two-center wavefunctions can be written
as combinations of target- and projectile-centered 1s-wavefunctions: $\Psi
_{\pm }=\left( \Phi _{p}\pm \Phi _{T}\right) /\sqrt{2}$.

Starting with a target localized electron we assign the initial conditions $%
a_{\pm }=1/\sqrt{2}$ at $t\longrightarrow -\infty $ and solve the equation $%
\left( {\rm \ref{coupled}}\right) $ numerically. Although at $%
t\longrightarrow -\infty $ the probabilities $\left| a_{\pm }\right| ^{2}$
remain very close to 1/2, the amplitudes $a_{\pm }$ acquire phases which
change the relative population of the projectile and the target 1s state. We
correct for energy conservation which feeds the increasing binding energy of
the electron back to an increasing relative motion energy of the two protons
as they come closer. This is specially important as $E_{p}$ becomes of order
of hundreds of eV, and smaller. In figure 1 we show the time dependence of $%
V_{\pm }\left( t\right) $ and $W\left( t\right) $ for $E_{p}=10$ keV and a
nearly central collision, $b=0.1$ \AA . One observes that the potentials $%
V_{\pm }\left( t\right) $ extend much farther out than $W\left( t\right) .$
Moreover, we find that as $E_{p}$ decreases the potential $W$ decreases
faster than the projectile's velocity, $v_{p}$. This is mainly due to the
derivative of \ in eq. (\ref{potent}). At $E_{p}\simeq 100$ eV the potential 
$W$ loses its relevance as compared to $V_{\pm }$, which have no dependence
on $v_{p}$. This becomes clear in figure 2. In this figure we show the
exchange probability as a function of the impact parameter for two 
projectile energies. 
The solid line
is the full solution of eq. $\left( {\rm \ref{coupled}}\right) .$ The
dashed line is the approximation obtained when we set $W=0$ in eq. $\left( 
{\rm \ref{coupled}}\right) .$ In the later case, the equations decouple and
it is straightforward to show that the exchange probability is given by

\begin{equation}
P_{exch}=\left| \sum_{\pm }a_{\pm }\left( \infty \right) \left\langle \Phi
_{T}|\Psi _{\pm }\left( \infty \right) \right\rangle \right| ^{2}=\frac{1}{2}%
+\frac{1}{2}\cos \left\{ \frac{1}{\hbar }\int_{-\infty }^{\infty }\left[
E_{-}\left( t\right) -E_{+}\left( t\right) \right] dt\right\} \;.
\label{approx}
\end{equation}
At $E_{p}=10$ keV there is an appreciable difference between the full
calculation and the approximation (\ref{approx}). But, for $E_{p}=100$ eV
the results are practically equal, except for very small impact parameters
at which the potential $W$ still has an effect.

One observes that the exchange probability is not constant at small impact
parameters, but oscillates wildly around 0.5, specially for low projectile
energies. One might naively assume that because the collision is almost
adiabatic, the system loses memory of to which nucleus the electron is bound
after the collision. Thus, for small impact parameters one would expect a
50\% probability of finding the electron in one of the nuclei at $t=\infty $%
. However, this is not what happens. From eq. (\ref{approx}) we see that 
minima of the probability occur for impact parameters satisfying the relation

\begin{equation}
\int_{-\infty }^{\infty }\left[ E_{-}\left( t\right) -E_{+}\left( t\right) %
\right] dt=2\pi \hbar \left( n+1/2\right) ,\;\ \ n=0,1,2,...,N.
\label{minima}
\end{equation}
This relation looks familiar, of course. It simply states that the
interference between the 1s$\sigma $ and the 2p$\sigma $ states induces
oscillations in the exchange probability. The electron tunnels back and
forth between the projectile and the target during the ingoing and the
outgoing part of the trajectory. When the interaction time is an exact
multiple of the oscillation time, a minimum in the exchange probability
occurs. The average probability over the smaller impact parameters is indeed
0.5. As the impact parameter decreases from infinity, the first maximum in
the exchange probability indicates the beginning of the region of strong
exchange probability. One sees that at low proton energies this starts at $%
b\simeq 3$ \AA . The size of the hydrogen atom is about 0.5 \AA\ and thus
the electron travels in a forbidden region (tunnels) of about 2 \AA\ from
the target to the projectile. This is possible because of the strong
interference between the 1s$\sigma $ and the 2p$\sigma $ states, which for
some trajectories satisfy the quantum relation (\ref{minima}).

To obtain the stopping power we need the total cross section for charge
exchange, $\sigma =2\pi \int P_{exch}bdb$. This is shown in figure 3. The
solid line is the full coupled-channels calculation, while the dashed line
uses approximation (\ref{approx}) for the exchange probability. We observe
that the approximation (\ref{approx}) reproduces well the full calculation
even at the highest energies. The reason is that the potential $W$ is always
smaller than $V_{\pm }$ for large impact parameters which have more weight
on the integral cross section. \ We also compare our calculations with the
lowest energy data of McClure \cite{Mc66}. The formalism developed here is
inappropriate for energies in the tens of keV range and higher, as the
projectile velocity becomes comparable to or higher than the electron
velocity. This implies that two-center states with higher energy and even
continuum states (ionization) should be included in the calculation. For $%
E_{p}\longrightarrow 0$, the charge exchange cross section becomes the
constant value $\sigma \left( E_{p}=0\right) =37.88$ $\times 10^{-16}$ cm$%
^{2}$. This happens because, when $E_{p}\longrightarrow 0$ and as the
projectile nears the targets, the increasing electron binding in the
two-center system  acts as a push in the relative motion energy to
compensate for energy conservation. The average result is that the cross
section for charge exchange becomes approximately constant for projectile
energies of tens of eV and below.

In figure 4 we show the stopping cross section of the proton. The stopping
cross section is defined as $S=\sum_{i}\Delta E_{i}\;\sigma _{i}$ , where $%
\Delta E_{i}$ is the energy loss of the projectile in a process denoted by $i
$. The stopping power, $S_{P}=dE/dx$, the energy loss per unit length of the
target material, is related to the stopping cross section by $S=S_{P}/N$,
where $N$ is the atomic density of the material. In our charge exchange
mechanism the electron is transferred to the ground state of the projectile
and the energy transfer is given by $\Delta E=m_{e}v_{p}^{2}/2$, where $v_{p}
$ is the projectile velocity. Assuming that there is a few free electrons in
the material (e.g., in a hydrogen gas) only one more stopping mechanism at
very low energies should be considered: the nuclear stopping power. This is
simply the elastic scattering of the projectile off the target nuclei. The
projectile energy is partially transferred to the recoil energy of the
target atom. The stopping cross section for this mechanism has been
extensively studied by Lindhard and collaborators (see, e.g., ref. \cite
{LSS63}). \  The nuclear stopping includes the effect of the electron
screening of the nuclear charges.

The dotted line in figure 4 gives the energy transfer by means of nuclear
stopping, while the solid line are our results for the charge-exchange
stopping mechanism. The data points are from the tabulation of Andersen and
Ziegler \cite{AZ77}. We see that the nuclear stopping dominates at the
lowest energies, while the charge-exchange stopping is larger for proton
energies greater than 200 eV. Since we neglect the difference between
molecular and atomic hydrogen targets, there is a limitation to compare our
results with the experimental data. But, the order of magnitude agreement is
very good in view of our simplifying assumptions. We do not consider the
change of the charge state of the protons as they penetrate the target
material. The exchange mechanism transforms the protons into H atoms. These
again interact with the target atoms. The can loose their electron again by
transfer to the 1s state of the target \cite{GS93}.

The best fit to our calculation for the stopping power for proton energies
in the range 100 eV - 1 keV yields $S\sim v_{p}^{1.35}$. This contrasts with
the extrapolation $S\sim v_{p}$, based on the Andersen-Ziegler table. But,
this discrepancy is much less than the one obtained by Golser and Semrad 
\cite{GS91} for helium targets, who found a stopping power for protons $%
S\sim v_{p}^{3.34}$ for protons in the energy range of 4 keV. No data at
lower energies are available in this case. But, the Golser and Semrad data,
for proton energies above 3 keV, firmly indicate that a high power
dependence on the projectile velocity will be also valid at lower energies,
in contrast the predictions from the Andersen and Ziegler tables \cite{AZ77}%
. One cannot extend our calculations to helium targets as the initial
wavefunction cannot be described in terms of a simple sum of two-center
states. A much larger two-center basis is necessary. Since the electrons in
the helium target are more bound than in the proton, the charge-exchange
probability must be much smaller than in the case of hydrogen targets. One
thus should indeed expect a much stronger dependence of the stopping on the
projectile velocity. At very low energies, of the order of some hundreds of
eV, the stopping cross section should be entirely dominated by nuclear
stopping, even more than for hydrogen targets.

The $p+p\longrightarrow d+e^{+}+\nu _{e}$ reaction is a very important one
occurring in, e.g., our sun. But, it proceeds via the weak interaction and
its cross section is extremely small for studies under the laboratory
conditions \cite{Cla68,RR88}. Fortunately, a good theoretical model exists
for this reaction \cite{Bah89}. Other reactions could be strongly influenced
by the stopping power of protons and deuterons due to the charge-exchange
mechanism. They can be relevant for the study of $d+D$ reactions in stellar
interiors and fusion reactors. Another application is the D(p, $\gamma $)$%
^{3}He$ reaction which is important for the hydrogen burning in stars. In
our sun the most effective energy of this reaction is  $E_{c.m.}=6.5\pm 3.3$
keV at $T=15\times 10^{6}$ K.\ At this energy one expects that the
charge-exchange stopping cross section should be as important as the
ionization cross section. Experimental data exist at the lowest energy value
of 16 keV \cite{GLS63,Bai70}. Although the extrapolation based on theory
appears to be under control in this case, it is worthwhile to consider a
better study of the stopping power for this reaction. The steep rise of the
fusion cross sections at astrophysical energies amplifies all effects
leading to a slight modification of the projectile energy \cite{BBH97}. Our
results show that the stopping mechanism does not follow a universal pattern
for all systems. This calls for improved theoretical studies of
charge-exchange effects and for their independent experimental verification.
\ \ \ \ \ \ \ \ \ \ \ \ \ \ \ \

{\bf Acknowledgments}
We would like to express our gratitude to profs. A.B. Balantekin, S.R. Souza
and L.F. Canto for useful comments and suggestions during the development of
this work. This work was partially supported by the Brazilian agencies:
CNPq, FAPERJ, FUJB, and by the MCT/ FINEP/CNPq(PRONEX) (contract
41.96.0886.00). (*) John Simon Guggenheim fellow.
\\ \ \ \ \ \ \ \ \ \ \ \ \ \ \

{\bf Figure Captions}

{\it Fig. 1} - Time dependence of the interaction potentials $%
V_{\pm }\left( t\right) $ and $W\left( t\right) $ for $E_{p}=10$ keV and a
nearly central collision, $b=0.1$ \AA . 

{\it Fig. 2} - The
exchange probability as a function of the impact parameter for two 
projectile energies. The solid line
is the full solution of eq. $\left( {\rm \ref{coupled}}\right) .$ The
dashed line is the approximation obtained when we set $W=0$ in eq. $\left( 
{\rm \ref{coupled}}\right) .$ 

{\it Fig. 3} - The
solid line is the full coupled-channels calculation for the charge-exchange 
cross section, while the dashed line
uses approximation (\ref{approx}) for the exchange probability. The 
experimental  data are from McClure \cite{Mc66}. 

{\it Fig. 4} - The stopping cross section of protons on H-targets. 
The dotted line in gives the energy transfer by means of nuclear
stopping, while the solid line are our results for the charge-exchange
stopping mechanism. The data points are from the tabulation of Andersen and
Ziegler \cite{AZ77}.

\ \ \ \ \ \\


\begin{thebibliography}{99}
\bibitem{Cla68}  D.D. Clayton, {\it Principles of Stellar Evolution and
Nucleosynthesis}, McGraw-Hill, New York, 1968

\bibitem{RR88}  C.. Rolfs and W.S. Rodney, {\it Cauldrons in the Cosmos},
Chicago Press, Chicago, 1988

\bibitem{ALR87}  H.J. Assenbaum, K. Langanke, and C. Rolfs, Z. Phys. {\bf %
A327}, 461 (1987)

\bibitem{SR95}  E. Somorjai and C. Rolfs, Nucl. Instum. Meth. {\bf B99}, 297
(1995)

\bibitem{SKLS93}  T.D. Shoppa, S.E. Koonin, K. Langanke, and R. Seki, Phys.
Rev. {\bf C48}, 837 (1993)

\bibitem{LSBR96}  K. Langanke, T.D. Shoppa, C.A. Barnes and C. Rolfs, Phys.
Lett. {\bf B369}, 211 (1996)

\bibitem{BFMH96}  J.M. Bang, L.S. Ferreira, E. Maglione, and J.M. Hansteen,
Phys. Rev. {\bf C53}, R18 (1996)

\bibitem{AZ77}  H. Andersen and J.F. Ziegler, {\it The stopping and ranges
of ions in matter}, Vol. 3, Pergamon press, New York (1977)

\bibitem{GS91}  R. Golser and D. Semrad, Phys. Rev. Lett. {\bf 14}, 1831
(1991)

\bibitem{GS93}  P.L. Grande and G. Schiwietz, Phys. Rev. {\bf A47} (1993)
1119; Phys. Rev.{\bf \ A58}, 3796 (1998); Nucl. Inst. Meth. {\bf B153}, 1
(1999)

\bibitem{Te30}  E. Teller, Z. Physik, {\bf 61}, 458 (1930)

\bibitem{Me97}  E. Merzbacher, {\it Quantum Mechanics}, John Wiley, New York
(1997)

\bibitem{Mc66}  G. W. McClure, Phys. Rev. {\bf 148}, 47 (1966)

\bibitem{LSS63}  J.Lindhard, M. Scharff and H.E. Shi\O tt, Mat. Fys. Medd.
Dan. Vid. Selsk. {\bf 33}, 1 (1963)

\bibitem{GLS63}  G.M. Griffiths, M. Lal and C.D. Scharfe, Can. J. Phys. {\bf %
41}, 724 (1963)

\bibitem{Bah89}  J. N. Bahcall, {\it Neutrino Astrophysics}, Cambridge
University Press, Cambridge, 1989

\bibitem{Bai70}  G.M. Bailey, G.M. Griffiths, M.A. Olivo, and R.L. Helms,
Can. J. Phys. {\bf 48}, 3059 (1970)

\bibitem{BBH97}  A. B. Balantekin, C. A. Bertulani, M. S. Hussein, Nucl.
Phys. {\bf A627}, 324 (1997)
\end{thebibliography}
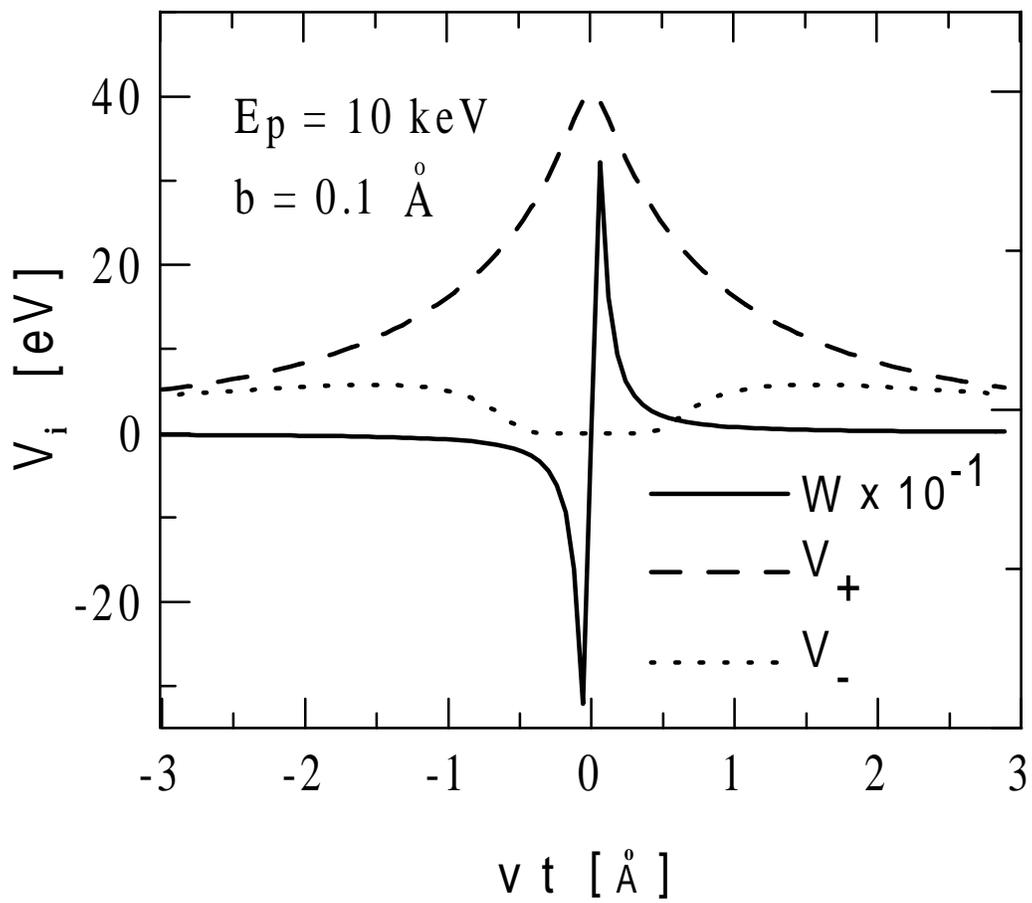
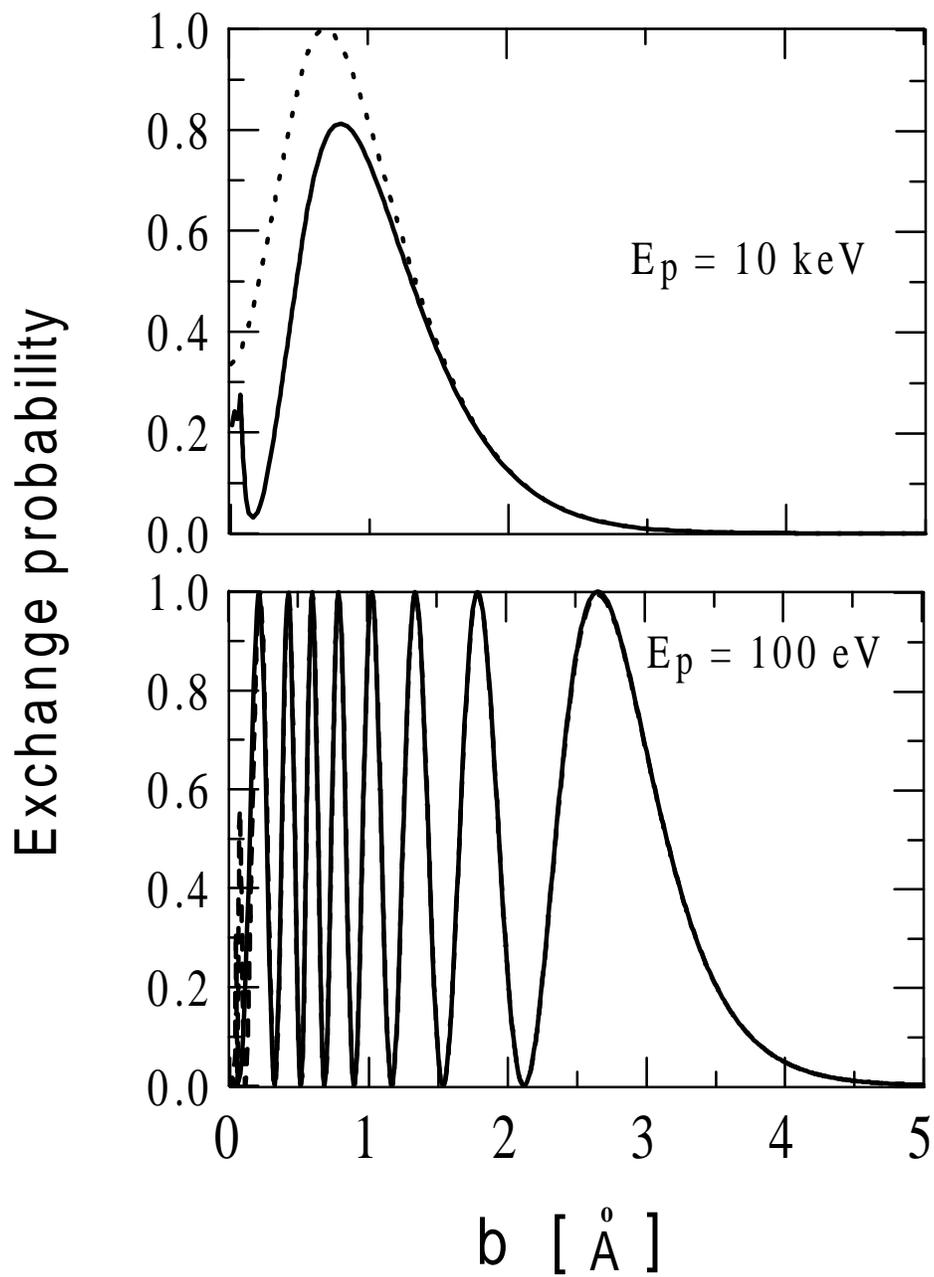
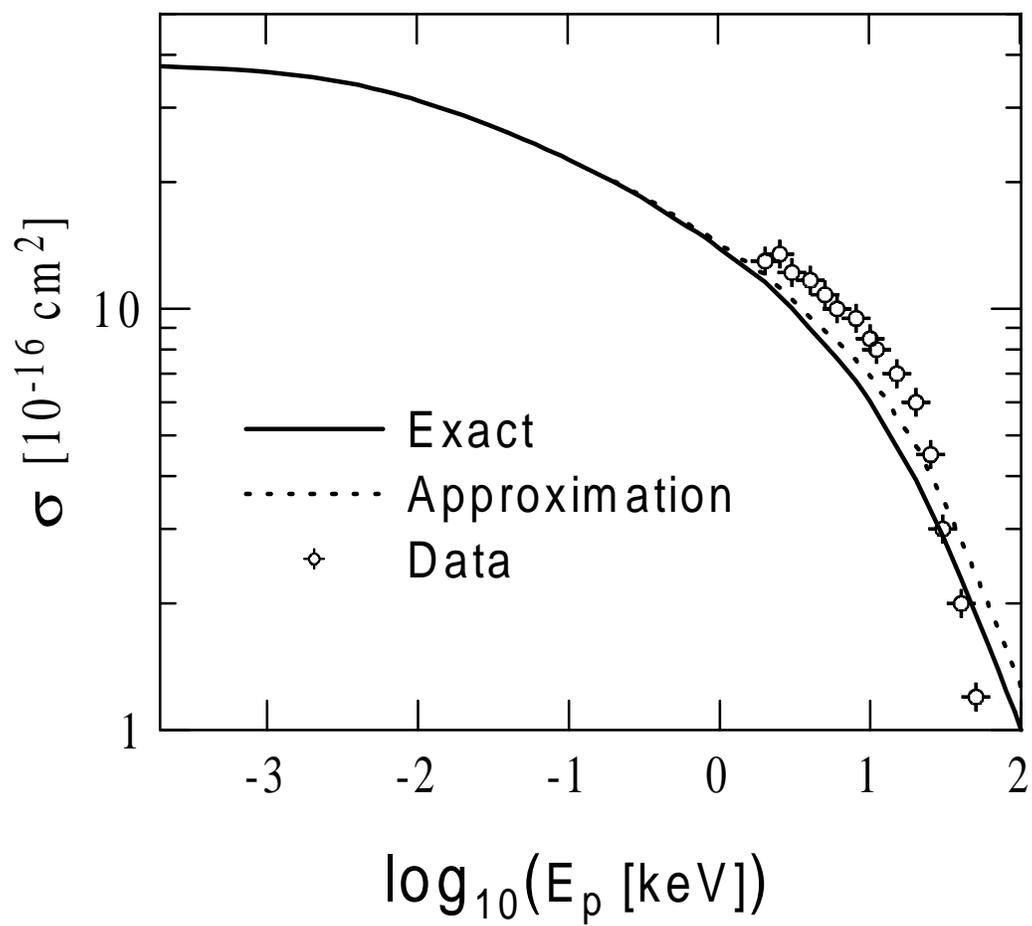
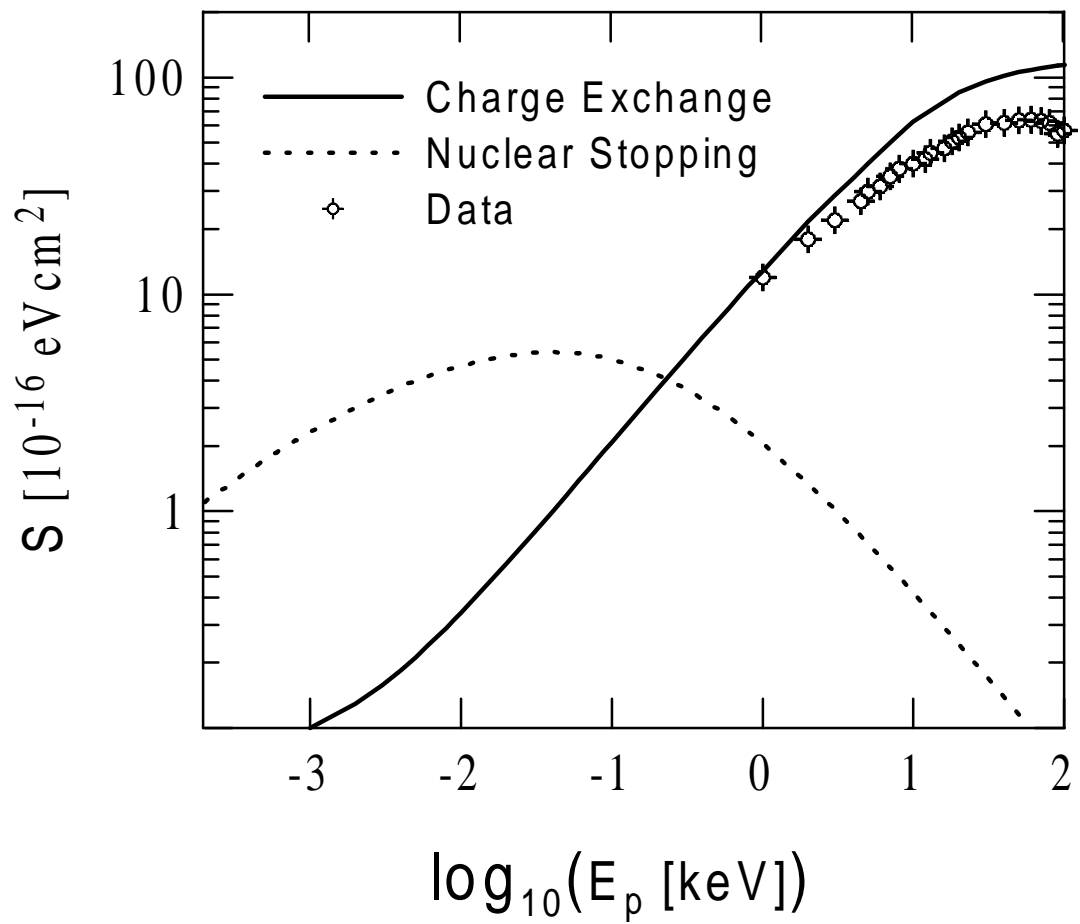
\end{document}